\documentclass{article}

\usepackage{arxiv}

\usepackage[utf8]{inputenc} 
\usepackage[T1]{fontenc}    
\usepackage{hyperref}       
\usepackage{url}            
\usepackage{booktabs}       
\usepackage{amsfonts}       
\usepackage{nicefrac}       
\usepackage{microtype}      
\usepackage{lipsum}
\usepackage{graphicx}
\usepackage{caption}
\usepackage{subcaption}
\usepackage{soul}
\usepackage[export]{adjustbox}
\graphicspath{ {./images/} }

\usepackage{color}

\title{Integrated Photonic Reservoir Computing with All-optical Readout}

\author{
 Chonghuai Ma \\
  \texttt{chonghuai.ma@ugent.be} \\
   \And
 Joris Van Kerrebrouck  \\
  \texttt{joris.vankerrebrouck@ugent.be} \\
  \And
 Hong Deng \\
  \texttt{hong.deng@ugent.be} \\
  \And
 Stijn Sackesyn  \\
  \texttt{stijn.sackesyn@ugent.be} \\
  \And
 Emmanuel Gooskens  \\
  \texttt{emmanuel.gooskens@ugent.be} \\
  \And
 Bing Bai  \\
  \texttt{baibing@photoncounts.com} \\
  \And
 Joni Dambre \\
  \texttt{joni.dambre@ugent.be} \\
  \And
 Peter Bienstman \\
  \texttt{peter.bienstman@ugent.be} \\
}

\begin{document}
\maketitle
\begin{abstract}
Integrated photonic reservoir computing has been demonstrated to be able to tackle different problems because of its neural network nature. A key advantage of photonic reservoir computing over other neuromorphic paradigms is its straightforward readout system, which facilitates both rapid training and robust, fabrication variation-insensitive photonic integrated hardware implementation for real-time processing. We present our recent development of a fully-optical, coherent photonic reservoir chip integrated with an optical readout system, capitalizing on these benefits. Alongside the integrated system, we also demonstrate a weight update strategy that is suitable for the integrated optical readout hardware. Using this online training scheme, we successfully solved 3-bit header recognition and delayed XOR tasks at 20 Gbps in real-time, all within the optical domain without excess delays. 
\end{abstract}


\section{Introduction}
The exponential growth of data in the digital age poses a significant challenge for data processing hardware units. As Moore's law approaches its limits, the traditional von Neumann hardware architecture also reveals its shortcomings when handling specific computational tasks, especially on machine learning related tasks. Concurrently, the medium for conveying information has shown signs of transitioning, with photonics emerging as a valid alternative to electronics in a lot of use cases, offering higher bandwidth, greater throughput, and lower power consumption. These advantages make photonics not only attractive for the communication industry, but also for information processing applications.

In this context, photonic reservoir computing (RC) \cite{Jaeger2004, Maass2002} has emerged as a promising computational architecture for optical information processing, offering next-generation high bandwidth, low latency, and low power consumption capabilities. Reservoir computing, a sub-paradigm of neural networks, is based on a recurrent neural network (RNN) system that is initialized randomly as a quasi-chaotic network (the reservoir). Training is delegated to a final linear readout layer. This modification yields two primary distinctions from other recurrent neural networks. First, the linear readout layer enables easier and more straightforward training. Second, the flexible requirements for the reservoir network allow for deployment on alternative hardware, shifting focus towards analog systems beyond traditional CPUs and GPUs.

Not surprisingly, reservoir computing is well-suited for implementation in photonic hardware, particularly integrated silicon photonics chips. Photonic reservoir computing combines high bandwidth and low power consumption signal processing in the optical domain, leveraging silicon photonics fabrication techniques that are also CMOS compatible, thus enabling low-cost, high-volume production capabilities.

Several hardware implementations of photonic reservoir computing exist, generally falling into two major categories: systems with a time-multiplexed single-node feedback loop and systems with spatially localized nodes. A comprehensive review of photonic reservoir computing strategies can be found in review papers \cite{lugnan2020photonic,shastri2020photonics}. Time-multiplexed reservoir computing systems with feedback loops are typically simpler and easier to implement, proving capable of solving various problems \cite{vinckier2015high, brunner2013parallel, dejonckheere2014all, nguimdo2014fast}. However, their application in real-time high-speed processing is limited due to data preparation requirements (e.g., signal masking). The system implemented in this paper belongs to the second category, featuring locally distributed nodes on a photonic integrated chip connected by simple optical waveguides. Hardware implementations of reservoir networks have been reported in \cite{Kristof14,katumba2019neuromorphic, van2017advances}. However, to achieve just-in-time high-speed photonic computation, both the reservoir network and the readout system need to operate in the optical domain \cite{Ma2021Bienstman-sra,Freiberger18} (details in {section2 }). With such all-optical integration, the computing unit can be considered an "optical-in, optical-out" system, introducing virtually no processing delay, except for the speed of light traveling in the optical waveguide.


The remainder of this paper is structured as follows. In {Section 2}, we describe reservoir computing in more detail and differentiate between electric and optical readout systems. We also briefly discuss leveraging computational nonlinearity in the optical readout, which is crucial for achieving real-time optical computation. In {section 3}, we present our hardware implementation of the system, discussing chip design and packaging. In {section 4}, we introduce our online training algorithm for the optical weights. Finally, We present the experimental results on a number of benchmarks in {section 5} and conclude our work in {section 6}.


\section{Photonic Reservoir Computing and Readout Architectures}

\begin{figure}[ht]
    \centering
    \includegraphics[width=1.0\textwidth]{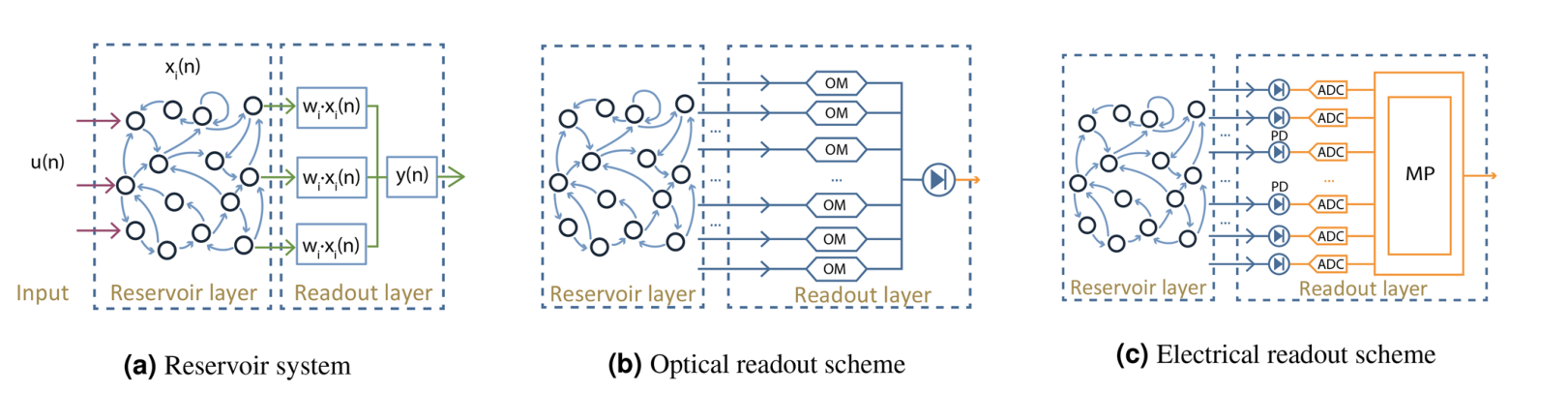}
    \caption{Reservoir readout systems (a) depicts the methematical schemetics of a reservoir system including both reservoir layer and readout layer. (b) and (c) illustrate two readout shemes. (b) employs an optical domain readout system and (c) utilizes an electrical readout system}
    \label{fig:res}
  \end{figure}

Reservoir computing is a subclass of recurrent neural networks (RNNs) that retains the primary advantage of RNNs in handling time-dynamic tasks due to the inherent feedback loops within the network. However, reservoir computing offers a more straightforward training process, making it an efficient alternative to traditional RNNs. Similar to RNNs, reservoir computing is adept at processing time-dynamic signals, but it differs in its architecture, comprising two major components (layers): the recurrent reservoir layer and the linear readout layer (Fig.~\ref{fig:res} a)). The key concept behind reservoir computing is that it does not depend on optimizing the internal interconnection parameters of the recurrent network, as is the case with RNNs. Instead, the reservoir layer remains unoptimized, while the readout layer applies a linear regression function to optimize the weights based on the outputs from the preceding reservoir layer.

By not needing to optimize the internal connection weights, the dependence of these weights on e.g. fabrication tolerances is mitigated to a certain extent. This is attractive for implementing an analog hardware system, as opposed to being restricted to digital CPU or GPU implementations. One approach involves the use of a photonic chip, which combines waveguides for signal connectivity and interference components that function as nodes. Random initialization of the network happens during fabrication, where different roughness values and other variations lead to interconnection weights with a random phase, which affects the interference at each node.

In a passive photonic reservoir network, the generic state update equation for discrete time can be written as:

$$x[k+1] = W_\mathrm{res}x[k] + w_\mathrm{in}u[k+1]$$

Here, $u$ represents the inputs of the reservoir, which are weighted by an input weights $w_\mathrm{in}$. For an $N$-node reservoir, the interval weight matrix $W_\mathrm{res}$ has dimensions $N \times N$. It is important to note that in a photonic reservoir system, all quantaties (both signals and weights) are complex valued.

In passive optical reservoir systems, the nonlinearity is provided by a photodiode, which converts a complex-valued optical field to a real-valued electrical power using a $| \cdot |^2$ operation. This happens in the readout, which is another crucial component of the system. In this paper, we differentiate between two types: electrical readout and optical readout. The primary distinction between them lies in whether or not the node signals are converted into electrical signals prior to the weighted-sum operation. As depicted in Fig.~\ref{fig:res} b), an optical readout applies weights directly to the coherent signal, with the final output being an optical superposition of all the weighted inputs. Conversely, Fig.~\ref{fig:res} c) shows an electrical readout scheme, which first detects and converts coherent signals from the reservoir, subsequently summing them in the digital electrical domain (either on-chip or off-chip).

In general, electrical readout systems can be perceived as more straightforward for measurements, because signal processing and learning can take place within the electrical domain. However, it is important to consider the potential costs associated with using an analog-to-digital converter (ADC) for each channel, which may impact overall affordability and energy-efficiency. Optical readout systems, on the other hand, require on-chip optical components such as Mach-Zehnder Interferometers (MZIs) or Microring resonators to apply the weights. However, optical readout systems offer minimal latency and power consumption, as they do not involve electro-optical or opto-electrical conversions, which tend to increase latency and consume more power. This feature enables numerous applications that necessitate a computing unit in an optical link, without incurring additional latency or power overhead. This paper focuses on the implementation of the integration of optical readout systems over electrical readout systems. For information on electrical readout systems, please refer to \cite{sackesyn2018enhanced}.

Furthermore, the two readout systems exhibit computational differences due to the position of the nonlinearity. The electrical readout system provides a nonlinear $| \cdot |^2$ operation on each channel, before summation, whereas in the optical readout system the nonlinearity is applied after the signals have been summed. A more in-depth exploration of the effect of this difference can be found in \cite{Ma2021Bienstman-sr}.

\section{Reservoir chip Design and Packaging}

\paragraph{Reservoir chip design}  

\begin{figure}[ht]
    \centering
    \includegraphics[width=0.7\textwidth]{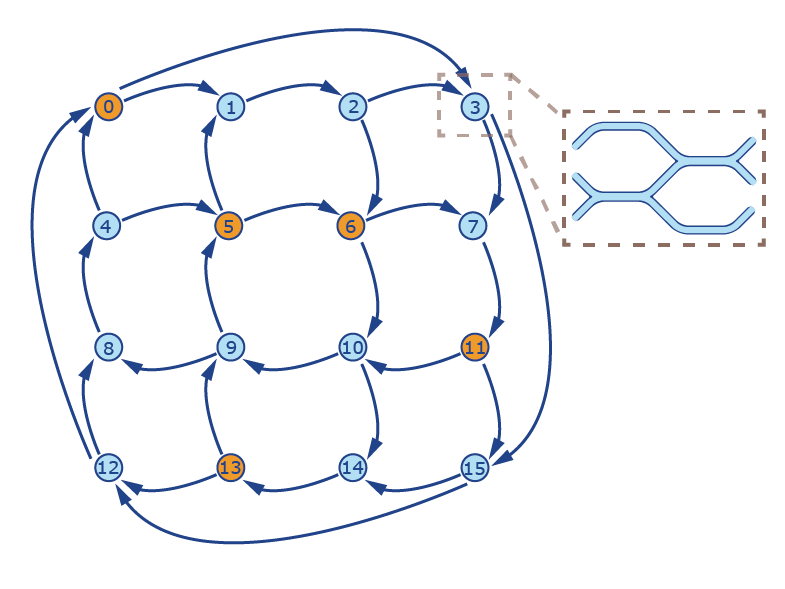}
    \caption{The 16-node reservoir network architecture. Each node is interconnected with four other nodes. Additionally, they all feature an extra input (which is exclusively used for nodes 0, 5, 6, 11, and 13 as indicated in orange) and an output. The 3x3 node is achieved by cascading two directional couplers, as demonstrated in the zoomed section at the top right}
    \label{fig:network}
  \end{figure}

For the reservoir chip, we employed a 16-node reservoir network with a 4-port architecture \cite{Sackesyn2021Bienstman-oeo}. This topology is designed to avoid radiation loss coming from combiners which have a smaller number of outputs than inputs. Each node is designed to handle three inputs (two from other nodes and one from the external input $u$) and three outputs (two to other nodes and one to the readout). A 3 x 3 Multimode Interferometer (MMI) could be utilized to establish this three-by-three connection. However, to mitigate losses resulting from fabrication deviations, we opted for a directional coupler (DC) as the interferometer unit, employing two DCs to create a three-by-three structure as the top right part of Fig.~\ref{fig:network}. This approach significantly reduces insertion loss. One drawback is that the splitting ratio of directional couplers is sensitive to fabrication tolerances, often leading to performance discrepancies between design and fabrication. Additionally, the splitting ratio can deviate locally even within the same die. Nevertheless, reservoir computing has been demonstrated to be tolerant of network interconnection initialization. Our simulations indicate that a maximum splitting ratio deviation of 30:70 should not significantly impact, rendering the DC-based low-loss node highly suitable for a photonic reservoir system.

In theory, our 16-node system can accommodate 16 inputs from external sources and 16 outputs to the readout layer. However, to maximize the mixing and richness of the recurrent network, we opted to input only five nodes and read all 16 nodes based on our simulation results. The input nodes selected are 0, 5, 6, 11, and 13, illustrated by Fig.~\ref{fig:network}.

The design of a photonic reservoir network must take into account the target operating signal speed. This is because, to achieve minimal noise and synchronized mixing between different signals, the time delay differences between two adjacent nodes should be on par with the time delay between neighboring bits of the operating signal. This time delay between two nodes is set by the waveguide length. In our work, the hardware was compatible with a speed of 20 Gbps,

After the reservoir, the 16 optical outputs are directed to the readout layer, where they are weighted and combined, with the resulting sum proceeding to a single optical output. As previously mentioned, we apply complex weights to coherent optical signals. We employ balanced Mach-Zehnder Interferometers (MZIs) as amplitude weighting elements and heater-based phase shifters to manipulate the phase of the optical signals.

\begin{figure}[h]
    \centering
    \includegraphics[width=0.8\textwidth]{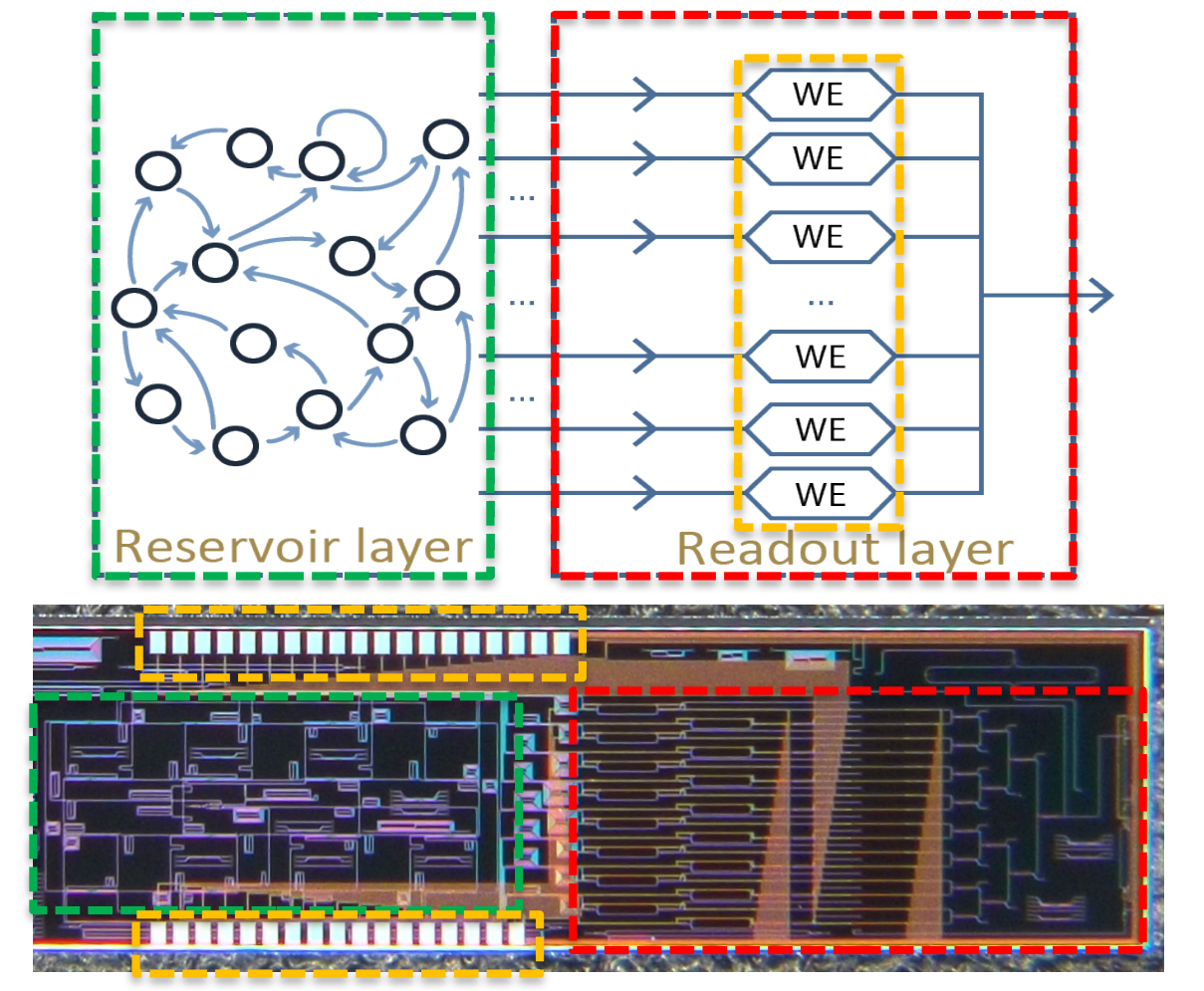}
    \caption{Chip layout. Our fabricated integrated photonics reservoir compututing system shown in the bottom. 
    The system is primarily composed of three parts: the reservoir layer (green box), the readout layer (red box), and the weight control. The I/O part of the weight control on chip is represented by the yellow box.}
    \label{fig:chip}
  \end{figure}

All of this is illustrated in Fig.~\ref{fig:chip}, where the left side of the chip area (enclosed by the green dotted frame) houses the reservoir network, while the right side (enclosed by the red dotted frame) accommodates the all-optical readout system.

The fabricated chip has dimensions of 3 mm by 10 mm and was produced using a 180 nm CMOS platform with a 220 nm silicon waveguide thickness and a 3 $\mu$m oxide thickness. 

\paragraph{Packaging}

The training process entails several iterations of heater voltage changes (resulting in temperature changes), signal acquisitions, and idle time for information processing. To ensure the chip functions correctly and produces accurate measurements at each training iteration, it is necessary to package it in a manner that safeguards it from environmental factors such as temperature fluctuations, mechanical stress, and air turbulence. In this work, we employ a fiber array to provide a stable and precise alignment between the chip and the optical fibers. 

We chose a 24-channel standard single-mode fiber (SMF) array for the right side of the chip. Due to the chip's geometry, we placed the input grating coupler on the left side of the chip. We used a single fiber and secured it to the input grating coupler with adhesive.

\begin{figure}[h]
    \centering
    \includegraphics[width=0.5\textwidth]{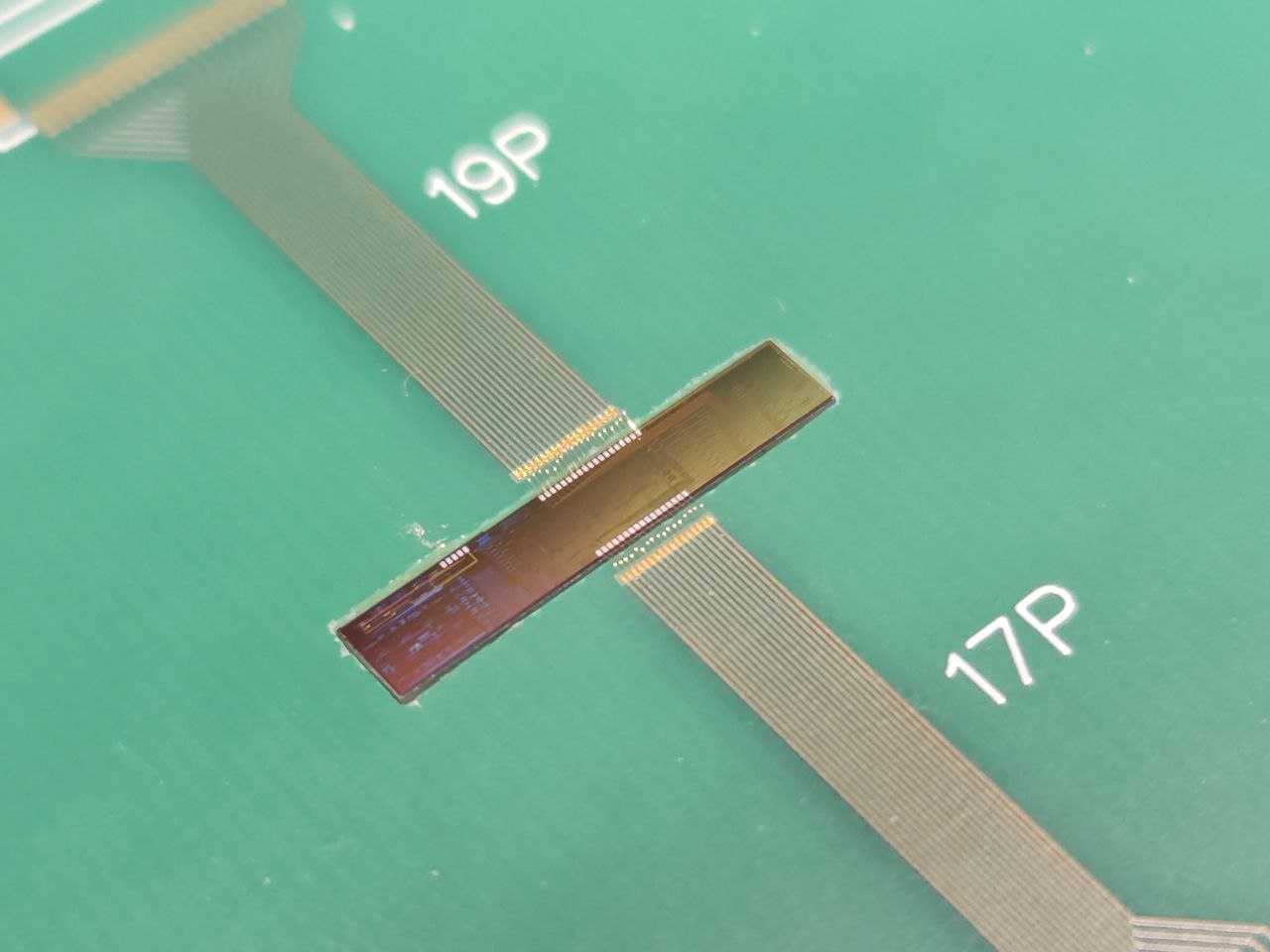}
    \caption{The chip is nestled within a hollowed-out region in the board, making a direct connection with the metal housing from its bottom side. All electronic input/output connections are DC. To efficiently utilize space, 19 contacts are on the top side and 17 on the bottom side of the chip.}
    \label{fig:wire-bond}
  \end{figure}

Additionally, we wire-bonded the chip to a designated printed circuit board (PCB) to ensure stable and convenient access to all on-chip phase shifters and amplitude controllers (Fig.~\ref{fig:wire-bond}). We've designed the Printed Circuit Board (PCB) with a hollowed-out structure at the center, specifically to house the chip. The bottom features an aluminum casing, enabling direct contact between the chip and the metal for efficient heat conduction.This arrangement allows for optimal and stable temperature control.

\section{Online Training of the Optical Readout}

In our design, the signal outputs from the reservoir network are directly injected into the readout system for weighting. This implies that the nodes are not inherently observable, as required for training the reservoir using a Moore-Penrose generalised matrix inverse \cite{penrose1955generalized}. 

We could have opted to fix this by also extracting the node signals before routing them to the readout system. Although this method is theoretically feasible, it presents several challenges in a practical implementation. First, we need both amplitude and phase information from each node of the network. While this can be achieved using a coherent receiver or a reference signal and an interferometer, it introduces additional noise and uncertainty, requiring a stable locking mechanism. The second challenge is the unknown phase response of the readout system itself. Assuming we successfully obtain all the amplitude and phase information from the output of the reservoir and then calculate the target weights from an offline learning process, it is still not guaranteed that we can program these weights in the readout system without errors. Indeed, the phase delay of the waveguide and Y-combiners is essentially random due to fabrication variation, and would require extensive calibration.

One could argue that the individual signal of each node could be observed at the final output of the readout system, including all the phase uncertainties of the readout, simply by setting its weight to unity and those of the other nodes to zero. However, this method still suffers from noise problems because it is impossible to completely "shut down" a channel due to the limited extension ratio of the modulators. 

Since all of the approaches above would require extensive calibration anyhow, we decided to forgo the temptation of trying to achieve complete node observability, and adopt black-box optimization as a more straightforward approach. As previously discussed, the reservoir computing architecture employs a linear regression in the readout layer, which implies that the optimization map is convex. For real-valued linear regression, a one-step solution like the Moore-Penrose generalized matrix inverse exists. However, for coherent photonic reservoir systems, the signals are represented by complex numbers. In this case, the pseudo-inverse solution still exists, but the solution requires a fixed complex number as the output signal. This means that the calculation solution implies a fixed phase of the optical signal before the photodetector, which is an unnecessary limitation. Upon detection, the phase of the signal launched on the detector does not affect the final readout. Therefore, although the pseudo-inverse works for a given complex-valued target, it fails to incorporate that the photodetector does not care about phase. This imposes unnecessary constraints on the optimisation. Nonetheless, the energy landscape is still convex, allowing us to utilize a relatively simple method, such as the random walks described in \cite{Bertsimas2004Vempala-ja}. However, some hardware limitations in practice still complicate the optimization and may even render it non-convex, posing challenges for implementing a pure random walk optimization algorithm. Therefore, we modify the optimization procedure to overcome these limitations. We will now discuss several aspects of this procedure, namely the optimisation procedure, the weight update strategy, and a so-called 'weight jumping' scheme. 

\paragraph{Optimization procedure}

For a certain iteration $n$, the optimization consists of the following steps:
\begin{enumerate}
\item Data acquisition. We obtain time traces from the real-time oscilloscope (RTO) and perform simple processing on the collected data, such as normalization and synchronization to align the output with the correct desired labels. For synchronization, we use a predefined bit pattern at the start of the sequence. Note that this procedure is is only necessary during the training phase.

\item Determining candidate weights. Based on the weight update strategy (discussed below), we choose which weights to update and select the size and the direction of the weight changes. This results in candidate weights $W_n'$.

\item Evaluation of performance of candidate weights. We calculate the L2 error between the processed real-time data $x_n$ and the target signal $\hat{y}$. If this error is lower than the error from the previous iteration, we adopt the candidate weights. Otherwise, the weights are unchanged with respect to the previous iteration.
\end{enumerate}

\paragraph{Weight Update Strategy} In every iteration, the weight update strategy determines which weights to train and how they are updated. If the weight change from the previous iteration resulted in a lower error, we keep updating the same set of weights in the same way. Otherwise, we randomly choose a number of weights to update. Each of these weights is randomly either increased or decreased with a given step size, which acts as a learning rate. At the begining of the optimization process, we typically start with a coarse optimisation involving a larger number of weights to update and a higher learning rate. As the procedure progresses, we opt for fewer updatable weights and a decreased learning rate (see section 5).

\begin{figure}[h]
     \centering
     \begin{subfigure}[b]{0.49\textwidth}
         \centering
         \includegraphics[width=\linewidth,right]{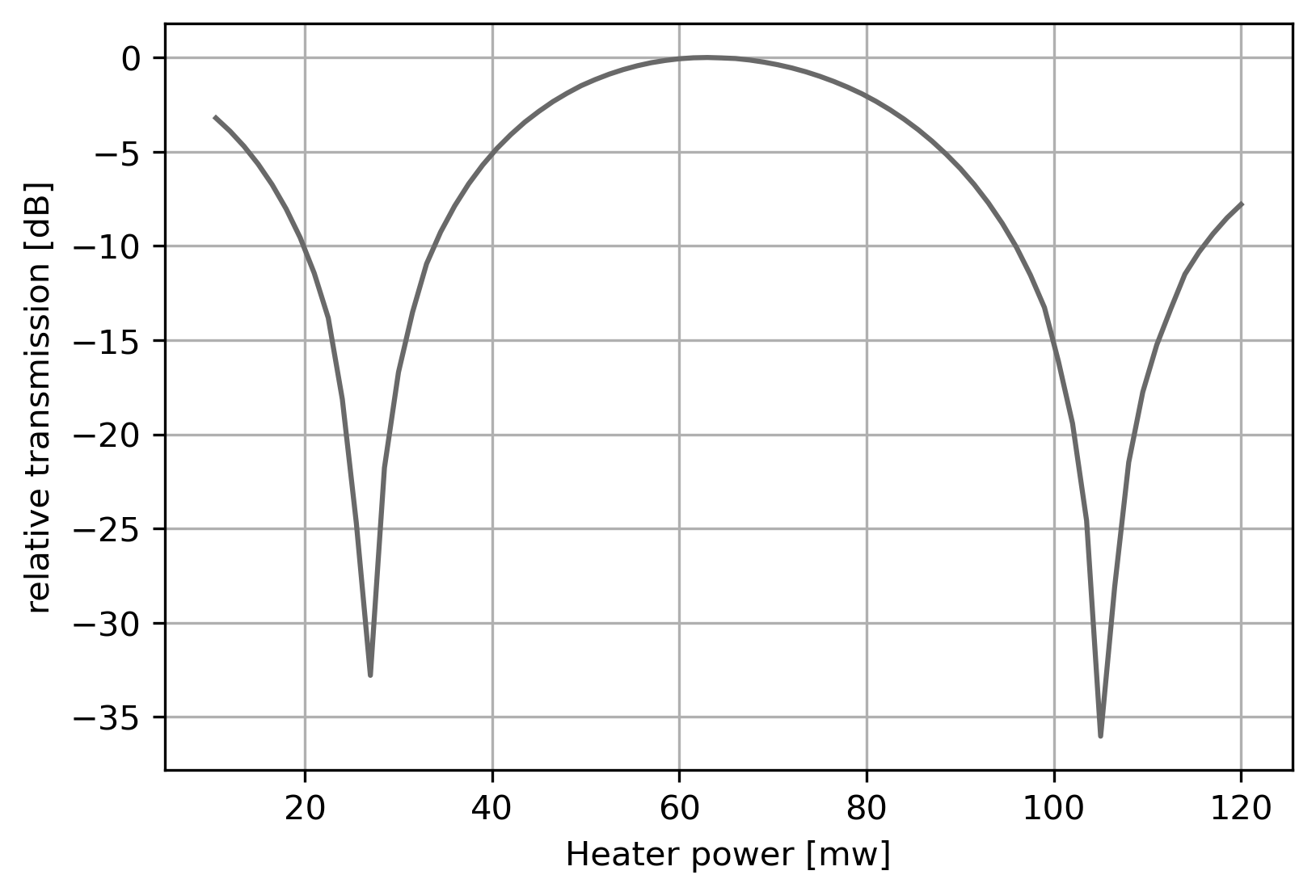}
         \caption{Example amplitude modulation as a function of heater power, The heater is embedded in one arm of a Mach-Zehnder Interferometer (MZI), which results in a periodic response.}
         \label{fig:heater_response}
     \end{subfigure}
     \hfill
     \begin{subfigure}[b]{0.49\textwidth}
         \centering
         \includegraphics[width=\linewidth, left]{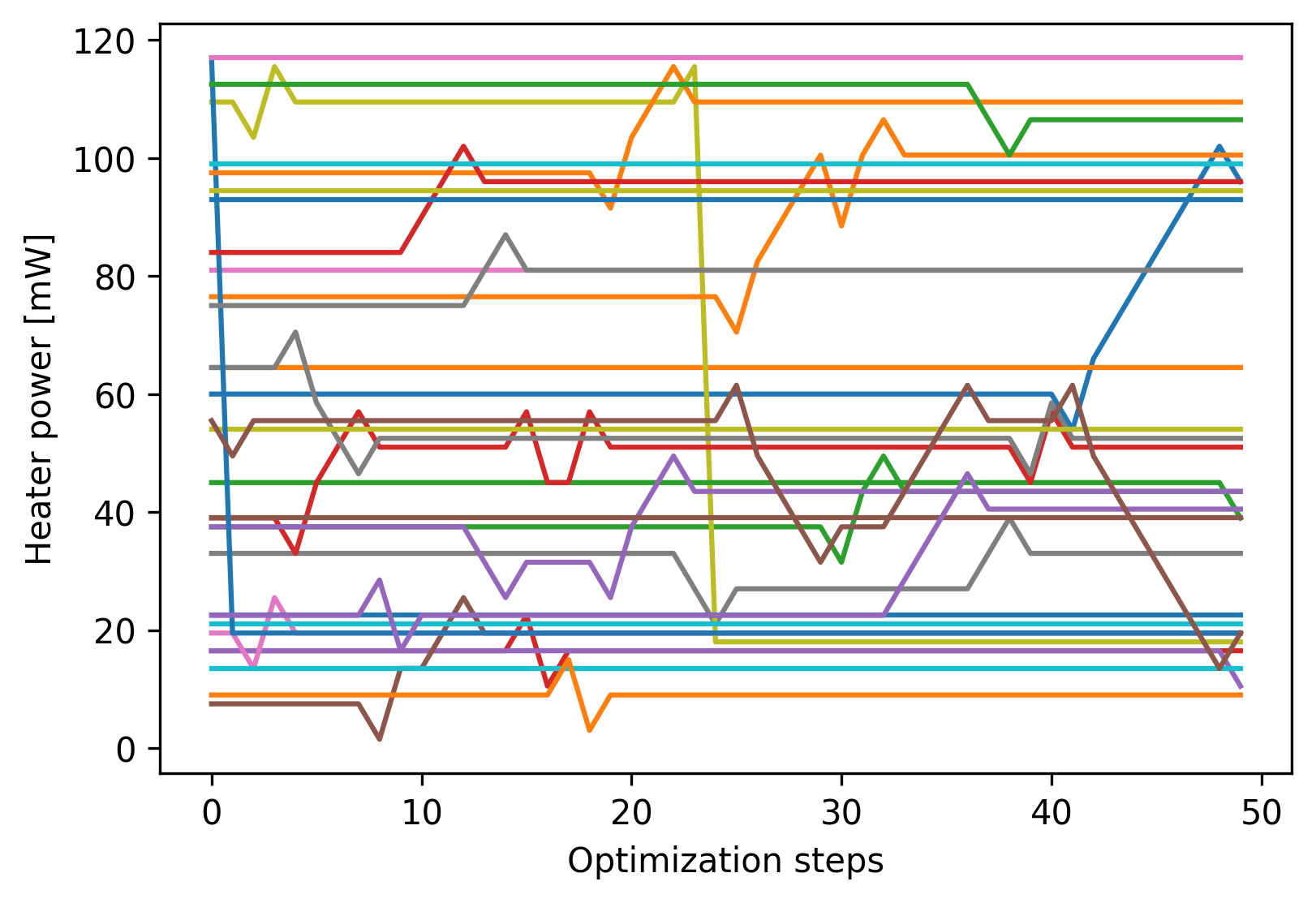}
         \caption{Heater power jump in a weight optmization session (spanning 50 iterations). The various colored lines correspond to the heater power levels for all the weights in the readout system.}
         \label{fig:weight_jump}
     \end{subfigure}
    \caption{Training}
    \label{fig:training}
\end{figure}

\paragraph{Heater Weight Jump} The complex-valued weights in our system are encoded as an amplitude modulation followed by a phase shift. The amplitude modulation uses a heater in one arm of a Mach-Zehnder Interferometer (MZI), while the phase modulation is done by a heater inside a single straight waveguide. In any case, the output of both of these elements is periodic as a function of heater current, as can be seen e.g. in Fig. \ref{fig:heater_response} for the amplitude modulation. Hovever, during optimisation it could happen e.g. that a current is updated from the upper end of one period to the lower end of the next period. While this is mathematically acceptable, we still choose to restrict the currents to a limited range of 1.5 to 2 periods. This is done in order to reduce the power consumption, which could also lead to thermal crosstalk. We do this by introducing a "weight jumping" scheme, whereby a current that is about to leave the range is reset to an arbitrary position at the other end of the range. Although this results in a disruptive change for the weights and the loss function, this is not a problem in the initial exploration phase. At the end of the learning process, where we take small steps, these jumps tend to become very rare anyhow.

As an example, Fig. \ref{fig:weight_jump} depicts a typical training session of 50 steps using the procedure introduced above. In each step, two random heater currents are updated. We can cleary observe the exploration going on for the different weights. At around the 24th step, one of the heater currents reached the predetermined upper bound and then was sent to a lower value, after which it continued to update.


\section{Experimental Results}

\paragraph{Measurement Setup}
The experimental measurements presented in this paper follow the all-optical readout scheme introduced earlier. The core concept involves maintaining an uninterrupted optical link connecting all key components without disruptions such as opto-electrical or electro-optical conversions. The optical signal source is generated by a continuous-wave (CW) laser (Santec TSL-510) operating at 1550 nm. The signal is then shaped by a high-speed lithium niobate electro-optic modulator (Photline MX-LN-40). Signal amplification before and after the reservoir computing chip is achieved using two erbium-doped fiber amplifiers (EDFAs). A high-speed photodiode (Finisar XPDV312R-VM-FA) is placed at the end of the optical link. In the figure, we omitted the polarization controllers, which are placed in the link before the modulator and the photonic chip.

\begin{figure}[h]
    \centering
    \includegraphics[width=0.5\textwidth]{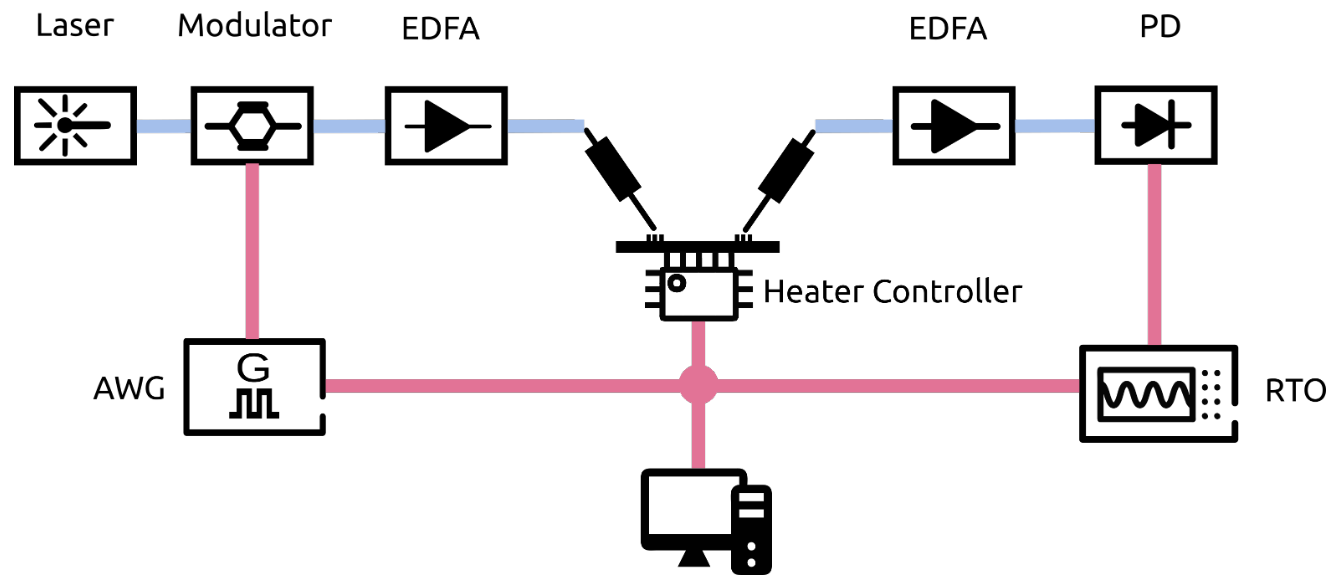}
    \caption{Schematic illustration of the setup used in the experiment. Optical connections are depicted by light blue lines, while electrical links are represented by light red lines. }
    \label{fig:setup}
  \end{figure}

In parallel with the optical link, electrical components are utilized. Signal definition in the electrical domain is achieved using an arbitrary waveform generator (AWG) from keysight(M8196A), which operates at a speed of 20 Gbit per second in this paper. The reservoir and readout system configurations on the chip are driven by voltage sources orchestrated by an in-house made digital-to-analog converter (DAC) control board that was made in house by our collaborators. The board provides a Python API for easy control by a computer. On the receiver side, we use a real-time oscilloscope (RTO) from keysight (DSAZ634A) to capture the waveform of the output signal from the reservoir chip. The signal is captured at each iteration and then fed to the training algorithm. All key components in the electrical link, including the AWG, the heater driver DAC, and the RTO, are controlled by a single computer, ensuring full control during the training process.

\paragraph{Header Recognition 3-bit}
We first tested our training process on a header recognition task, where the system should output a high value only if it recognises a certain 3-bit header (110 in this case). This task requires network memory, but does not demand significant computational nonlinearity. The training set for the header recognition task consists of 16,384 ($2^{14}$) randomly generated bits, with 1,024 bits reserved for a pre-specified sequence to facilitate better signal synchronization during training. This synchronization portion of the bitstream is only needed during training and is not required for validation and normal operation. 

\begin{figure}[h]
     \centering
     \begin{subfigure}[b]{0.49\textwidth}
         \centering
         \includegraphics[width=\linewidth,right]{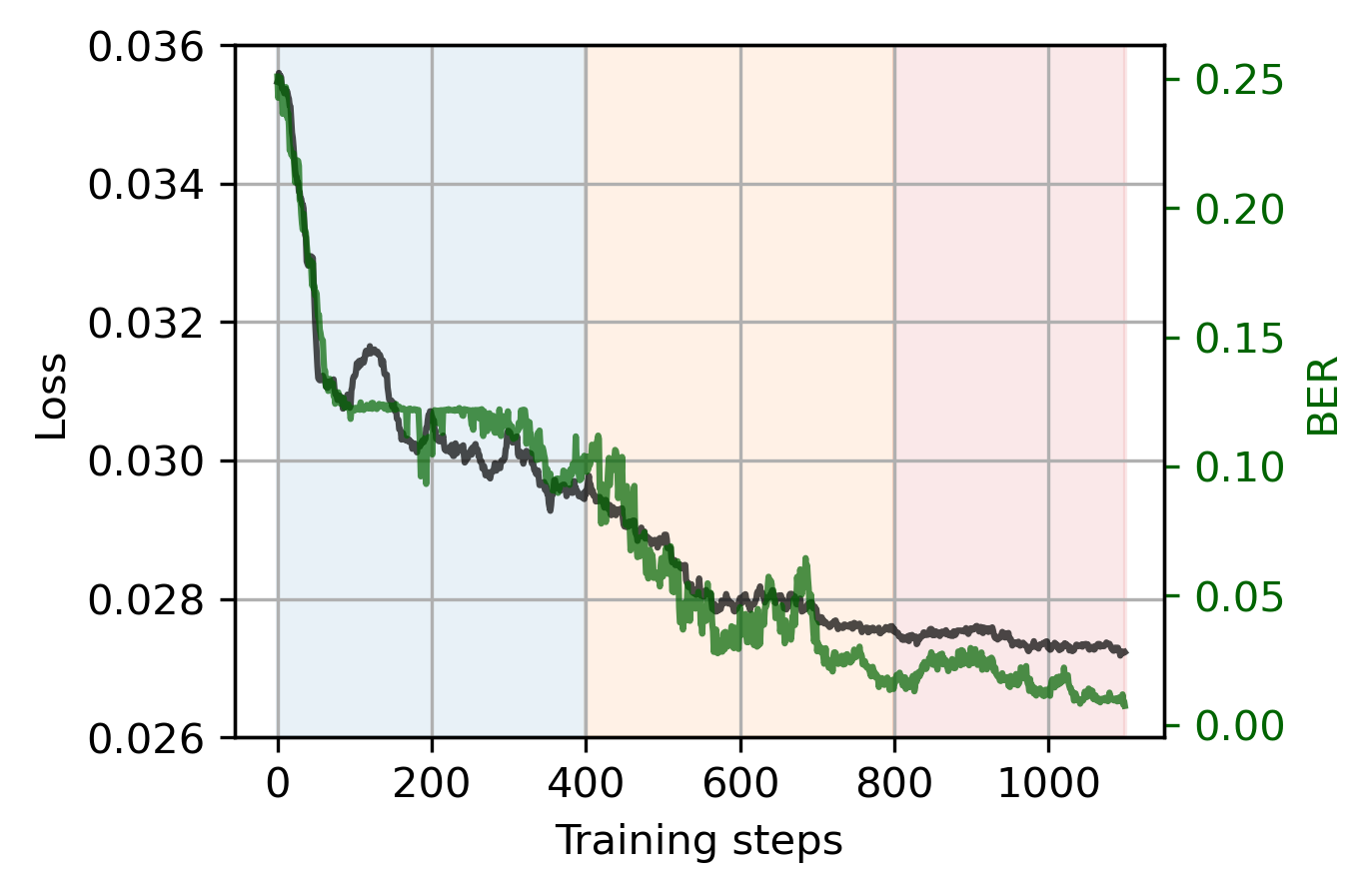}
         \caption{}
         \label{fig:hr_training1}
     \end{subfigure}
     \hfill
     \begin{subfigure}[b]{0.49\textwidth}
         \centering
         \includegraphics[width=\linewidth, left]{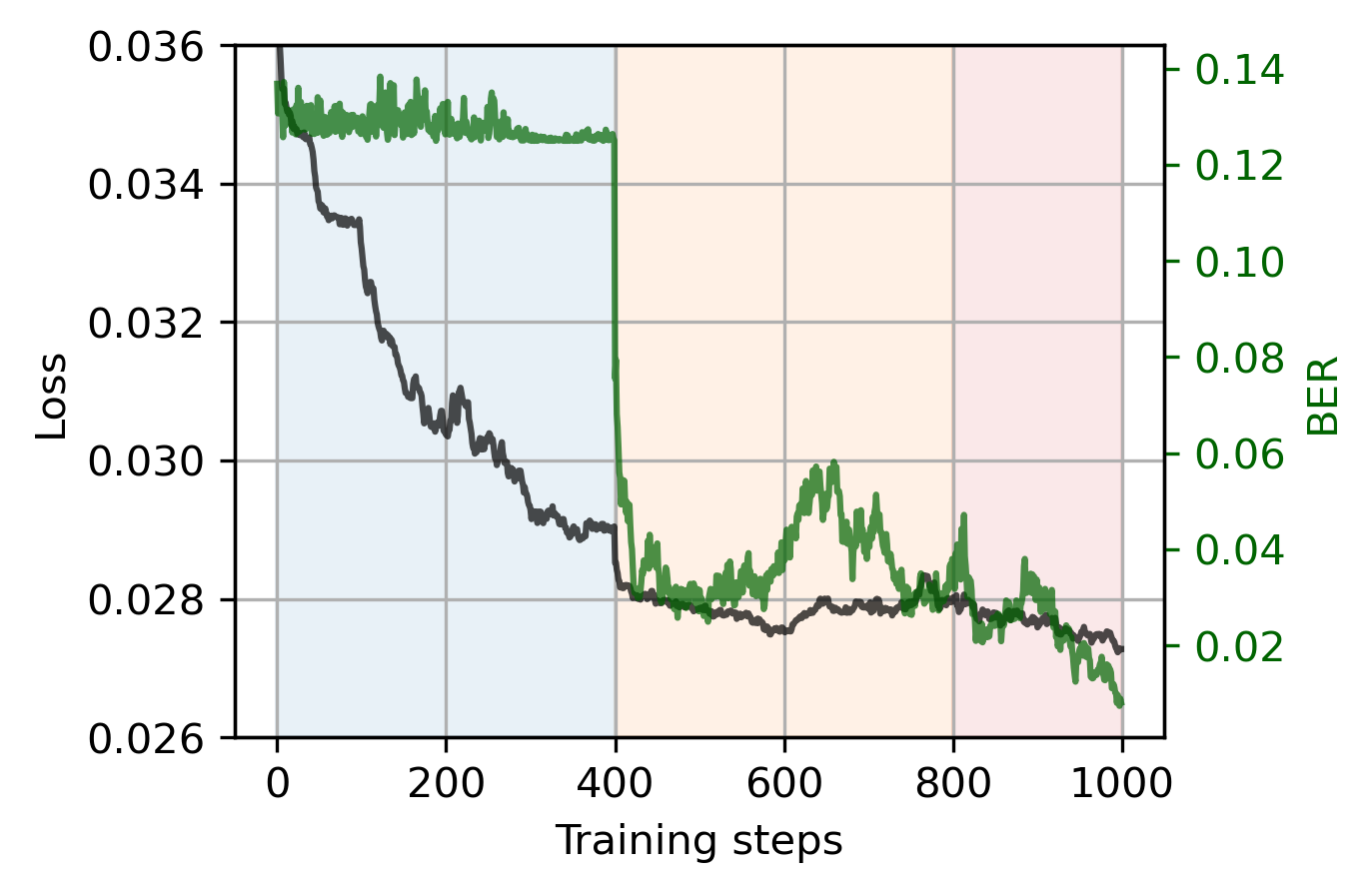}
         \caption{}
         \label{fig:hr_training2}
     \end{subfigure}
    \caption{header recognition training processes of two individual optimizations.The dark and green lines correspond to the L2 loss value and the Bit Error Rate (BER), respectively. Different training stages, with finely tuned optimization options (such as step size and number of weights to be optimized per iteration), are color-coded by vertical spans on the time axis.}
    \label{fig:hr_training}
\end{figure}

Fig. \ref{fig:hr_training} presents two individual training processes we performed, with real-time mean square error (MSE) and bit error rate (BER). For both cases, the optimization was carried out in three stages, corresponding to different training hyperparameters. As training begins, it optimizes three different weights per iteration, with a large step size of 3 minimum resolution (4.5 mW) for 400 iterations. As the training progresses, we limited the number of trainable weights to two and then one, with a smaller training step size of 3 mW and 1.5 mW (minimum power steping size) for another 400 iterations. The first training session demonstrates good alignment between the MSE and the BER. The color-coded vertical spans mark the training stages with different fine-tuned training parameters.

The second training session (illustrated by Fig\ref{fig:hr_training2}) shows a different optimization path. Note that here, the BER and MSE results are not well-aligned during the first stage of coarse optimization. This is because the real-time calculation of BER is unable to capture the performance improvement at the initial stage. In this specific region, the threshold setting of the BER calculation procedure might get trapped in a local minimum. This is possible as it is the initial phase of the optimization, and the bit stream may not yet be adequately shaped to provide a clear BER reading. This is also why we always choose to use MSE as the figure of merit during optimization. As this example shows, during training, BER is neither sensitive nor accurate enough to represent the training progress. However, in both training procedures, an MSE of 0.028 and a BER of around 0.01 are reached with around 1,000 iterations. This demonstrates the effectiveness and reproducibility of our optimization scheme, regardless of whether the BER manages to capture the training performance.

\paragraph{XOR Task}
Next we try the 1-bit delayed exclusive OR (XOR) task, a typical test for reservoir systems. which requires greater computational capability in terms of  the computational nonlinearity than the header recognition task. The objective of this task is to compute the XOR operation between the current bit and the one from a single bit period earlier. The XOR gate stands out among elementary logic gates due to the fact that its truth table is not linearly separable. In order to effectively train the photonic neuromorphic system, we again utilized a data stream comprising 16384($2^{14}$) bits. The experiment was conducted at a bitrate of 20 Gbit/s, while the continuous-wave (CW) laser operated at a wavelength of 1550 nm. Owing to the implementation of the optical readout scheme, the system operates entirely within the optical domain, thereby eliminating the need for opto-electrical or electro-optical conversions in the signal link. As a result, the overall throughput is not constrained by the reservoir system. The system experiences only a minimal time delay, primarily attributed to the optical path within the architecture, which is approximately 1 ns. 

\begin{figure}[h]
    \centering
    \includegraphics[width=0.8\textwidth]{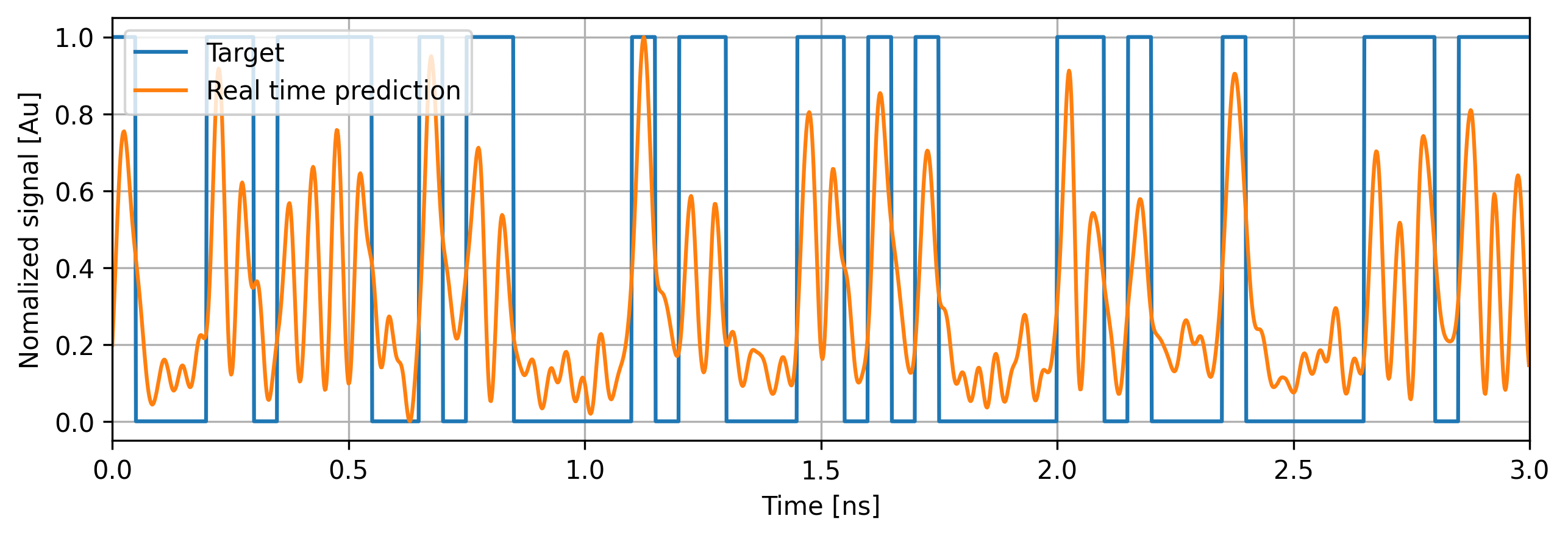}
    \caption{Optmized real-time prediction signal for the XOR task. The orange line illustrates the prediction signal, while the blue line depicts the target signals used during training.}
    \label{fig:xor_stream}
  \end{figure}
\begin{figure}[h]
    \centering
    \includegraphics[width=0.8\textwidth]{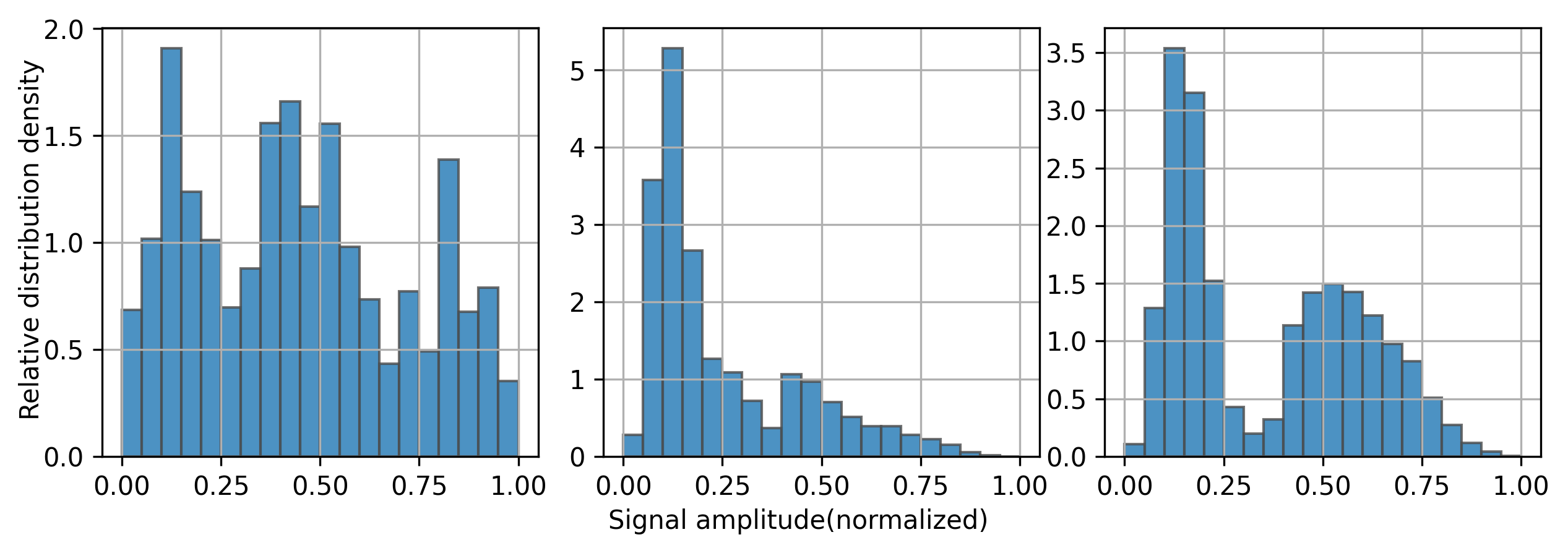}
    \caption{XOR task prediction histogram at different stages of the training proceedure. Initially, the ones and zeros are not distinctly separated. As the training progresses, the predicted ones and zeros become increasingly differentiated.}
    \label{fig:xor_histogram}
  \end{figure}

Fig. \ref{fig:xor_stream} presents the performance of our photonic circuit in executing this 1-bit delayed XOR task, showcasing real-time prediction in comparison to the target output. It can be observed that the real-time prediction signal consistently tracks the target signal, exhibiting a distinct contrast between binary 1s and 0s, which allows for the straightforward selection of a threshold value. By adopting a threshold of 0.32, we achieved a bit error rate (BER) of $4\times10^{-3}$. 
Figure \ref{fig:xor_histogram} provides a comprehensive visualization of the training progression in solving the task, represented by histograms of bit predictions at three distinct stages: before training, midway through training, and at the completion of training. By analyzing these histograms, we can gain a deeper statistical insight into the evolving distribution of target predictions throughout the training process. In the initial phase, when the optical network is untrained, the predictions are randomly distributed, closely resembling a uniform distribution. This random scattering is expected, as the network has not yet learned to discern any underlying patterns or relationships. As the training advances to the midway point, the network starts to distinguish between ones and zeros, although it is still not adept at predicting the majority of bits accurately. This stage is characterized by most bits being predicted within the lower energy region, a consequence of the dominant loss function contribution yielding lower optical output when the network is not fully trained. As the network optimization proceeds further, its performance improves, and an increasing number of ones are successfully predicted. The histogram begins to exhibit a balanced count between zeros and ones, indicating a more accurate representation of the XOR task. As a result of the training process, a discernible threshold naturally emerges, clearly illustrating the network's enhanced capability in separating ones and zeros.

\section{Conclusion}

In this work, we have presented a comprehensive study of an packaged  photonic reservoir computing system with an all-optical readout, demonstrating its capabilities in addressing bit-level tasks. The system leverages the intrinsic benefits of photonic platforms, including high-speed operation, minimal time delay, and reduced power consumption. By preserving signal processing exclusively within the optical domain, it effectively circumvents the constraints imposed by opto-electrical or electro-optical conversions.

A key aspect of our approach is the implementation of a flexible and adaptive weight update strategy, which significantly contributes to the enhanced performance of the system in terms of speed and effectiveness during the training phase. This optimization scheme also allows for adjustments in the middle of the optimization process, providing adaptability that translates to improved outcomes.

The experimental results showcased in this study indicate the system's ability to solve the header recognition task and the XOR task with low error rates. The real-time prediction signals exhibited a strong correlation with the target signals, with a clear threshold selection. The training progression also provided insight into the evolving distribution of target predictions, as the network learned to discern underlying patterns and relationships. The high-speed information processing capabilities of the system, operating at 20 Gbps bit rate, hold significant potential for various applications requiring real-time data analysis and processing. It's worth noting that the approach is not limited to a specific bit rate. By utilizing shorter delay lines, we can scale the bit rate up to several GHz. The only limiting factor comes from the corresponding electrical equipment. This performance underscores the advantages of photonic reservoir computing systems for tackling complex computational tasks in high-speed links.

\section{Acknowledgements}
This work was performed in the context of the Flemish FWO project G006020N and the Belgian EOS project G0H1422N.

\bibliographystyle{unsrt}  
\bibliography{references}  




\end{document}